\magnification=1095
\hsize=168 true mm
\vsize=230 true mm
\font\title= cmr12
\font\small= cmr8
\font\smallit = cmti8
\font\smallbf = cmbx8
\def\tr{{\rm tr}}
\def\next{\hfil\break\noindent}
\def\d{\partial}
\def\f#1#2{{\textstyle{#1\over #2}}}

\def\next{\hfil\break\noindent}

\def\Quadrat#1#2{{\vcenter{\hrule height #2
  \hbox{\vrule width #2 height #1 \kern#1
    \vrule width #2}
  \hrule height #2}}}
\def\dAlembert{\mathop{\kern 1pt\hbox{$\Quadrat{6pt}{0.4pt}$} \kern1pt}}

{\ }
\vskip 20 true mm

{\title 
\centerline{EXISTENCE AND NON-EXISTENCE RESULTS FOR} 
\centerline{GLOBAL CONSTANT MEAN CURVATURE FOLIATIONS}}

\vskip 14pt
\centerline{ALAN D. RENDALL}

{\small
\centerline{Max-Planck-Institut f\"ur Gravitationsphysik, Schlaatzweg 1,
14473 Potsdam, Germany}}

\vskip 9pt
{\leftskip 9 true mm
\rightskip 9 true mm
{\smallit Key words and phrases:}
{\small CMC hypersurfaces, Lorentz manifolds, Einstein's equations}}

\vskip 15 true pt

\centerline{
{\small 1. INTRODUCTION}}

\vskip 11 true pt 

If $(M,g)$ is a Lorentz manifold and $S$ a spacelike hypersurface, let $h$ and 
$k$ denote the induced metric and second fundamental form of $S$ respectively.
The mean curvature of $S$ is the trace $\tr_h k$. An interesting class of 
spacelike hypersurfaces are those whose mean curvature is constant (CMC
hypersurfaces). The Lorentz manifolds of primary interest in the following are
those which possess a compact Cauchy hypersurface and whose Ricci tensor 
satisfies $r(V,V)\ge 0$ for any timelike vector $V$. They will be referred 
to as cosmological spacetimes. The curvature condition is known as the strong 
energy condition and implies certain uniqueness statements for CMC 
hypersurfaces. Under very general conditions, if a region of a cosmological
spacetime is foliated by compact CMC hypersurfaces, then each of these has
a different value of the mean curvature and a time coordinate $t$ can be
defined by the condition that its value at a point be equal to the mean
curvature of the leaf of the foliation passing through that point. A time
coordinate of this type will be referred to in the following as a CMC time
coordinate.

The questions, whether a cosmological spacetime contains a compact CMC 
hypersurface and which values the mean curvature can take on a hypersurface of
this kind may seem purely geometrical in nature. However it turns out that the
answers to these questions depend crucially on factors which have no obvious 
geometrical interpretation, but which have a physical meaning, when the 
Lorentz manifold is considered as a model for spacetime. 

The Einstein equations for a Lorentz metric $g$ take the form $G=8\pi T$,
where $G$ is the Einstein tensor of the metric $g$ and $T$ is the 
energy-momentum tensor. To get a determined system of evolution equations
for the geometry and the matter, it is necessary to say more about the
nature of the matter model used. This means specifying some matter fields,
denoted collectively by $\phi$, a definition of $T$ in terms of $g$ and $\phi$
and the differential equations which describe the dynamics of the matter.
Putting these things together with the Einstein equations gives a system
of evolution equations, the Einstein-matter equations. It will be seen 
that the existence of global foliations by CMC hypersurfaces in a
cosmological spacetime which is a solution of the Einstein-matter equations
depends essentially on the matter model chosen.

Let $s$ denote the scalar curvature of the metric $g$. The
spacetime is said to satisfy the weak energy condition if 
$r(V,V)\ge (1/2)sg(V,V)$ for all timelike vectors $V$. (Note
that, despite the terminology, the strong energy condition does not imply
the weak one.) There is a topological obstruction to the existence of a 
compact spacelike hypersurface with vanishing mean curvature (maximal
hypersurface) in a spacetime which satisfies the weak energy condition. In a 
cosmological spacetime this implies that if the Cauchy hypersurface is a 
manifold which admits no Riemannian metric with non-negative scalar curvature
then the spacetime contains no compact maximal hypersurfaces. Moreover, there 
are strong restrictions in the case of a manifold which admits no Riemannian
metric of positive scalar curvature. The main questions are whether a
cosmological spacetime satisfying the weak energy condition and containing
at least one compact CMC hypersurface can be covered by 
a foliation of compact CMC hypersurfaces and whether the mean curvature of
these hypersurfaces takes on all values not forbidden by the topological
obstruction already mentioned. In other words, does it take all values in
the interval $(-\infty,0)$ or $(0,\infty)$ in the case where there is an
obstruction and all real values in the case there is none? This statement 
concerning the values attained by the mean curvature is equivalent to a
statement of whether a solution of the Einstein-matter equations exists
globally in a CMC time coordinate. The theorem proved in Section 2 shows by 
example that the answer is negative if no restriction is put on the matter 
model used. In Section 3 results are reviewed which show that under 
certain symmetry assumptions there are matter models for which the answer is 
positive. Possible extensions of these results are also discussed. For 
general information on the points just mentioned, the reader is referred to 
[1], which is complementary to the treatment here.

These results are related to the concept of crushing singularities. A
cosmological spacetime is said to have an initial crushing singularity
if there is a compact spacelike hypersurface $S_{t_0}$ and a foliation of the 
past of $S_{t_0}$ by compact spacelike hypersurfaces $S_t$, such that as $t$
tends to its limiting value towards the past, the mean curvature of $S_t$
tends uniformly to $-\infty$. (There is a similar definition with 
\lq initial\rq\ replaced by \lq final\rq , \lq past\rq\ replaced by 
\lq future\rq\ and $-\infty$ replaced by $\infty$.) If the $S_t$ form a
CMC foliation and the parameter $t$ labelling the leaves of the foliation
is a CMC time coordinate, then a sufficient condition in order to have an
initial crushing singularity is that $t$ should take all values in the 
interval $(-\infty,t_0)$. The positive results of Section 3 prove that 
this is true under certain hypotheses on the matter model and the symmetry
of the spacetimes considered, and so prove the existence of crushing 
singularities under certain circumstances. On the other hand, the results
of Section 2 provide examples of spacetimes where an initial (or final)
crushing singularity is not present. Informally, this can be expressed
by saying that the initial (or final) singularity in these spacetimes is
not crushing. It will now be indicated briefly how the statement about
the absence of crushing singularities follows from the theorem proved
in Section 2. If there is an initial crushing singularity, the hypersurfaces
$S_t$ provide barriers, as they are used in the well known existence theory 
for CMC hypersurfaces. Thus it can be assumed without loss of generality
that the $S_t$ are in fact CMC hypersurfaces and that $t$ is a CMC time
coordinate. Combining this with the uniqueness theorems for CMC 
hypersurfaces would show that the local CMC foliation which exists close
to the initial hypersurface could be extended to arbitrarily negative
values of $t$, contradicting the conclusions of the theorem.    

The techniques which are used to obtain the positive and negative results
are closely related. They will be explained in the case of spacetimes
with $U(1)\times U(1)$ symmetry where the Cauchy hypersurface has the
topology of a three-dimensional torus. Assume that a solution of the 
Einstein-matter equations satisfying the strong energy condition with a CMC 
Cauchy hypersurface of topology $T^3$ is invariant under the action of the 
group $U(1)\times U(1)$ consisting in rotating two of the three $S^1$ factors.
Then (see [2]) in a neighbourhood of the initial hypersurface the metric can 
be written in the form:
$$-\alpha^2 dt^2+A^2[(dx+\beta^1 dt)^2+
a^2\tilde g_{AB}(dy^A+\beta^A dt)(dy^B+\beta^B dt)]\eqno(1.1)$$
where $t$ is a CMC time coordinate. Here the coordinates are $t,x,y^2,y^3$. 
Upper case Roman indices take the
values $2,3$ while lower case ones take the values $1,2,3$, the value $1$
corresponding to the coordinate $x$. The functions $\alpha$, $\beta^a$,
$A$ and $\tilde g_{AB}$ depend on $t$ and $x$ and $\tilde g_{AB}$ has unit
determinant. They are periodic in $x$. The quantity $a$ depends only on
$t$. Some of the field equations are: 
$$\eqalignno{
&\d_x^2(A^{1/2})=-\f18A^{5/2}[\f32(K_1-\f13t)^2-\f23t^2
+2\eta_A\eta^A+\tilde\kappa^{AB}\tilde\kappa_{AB}+\tilde\lambda^{AB}
\tilde\lambda_{AB}
+16\pi\rho]&(1.2)                                         \cr
&\d_x^2\alpha+A^{-1}\d_x A\d_x \alpha=\alpha A^2[\f32(K_1-\f13t)^2+\f13 t^2
&\cr
&\qquad+2\eta_A\eta^A+\tilde\kappa_{AB}\tilde\kappa^{AB}
+4\pi(\rho+\tr S)]-A^2&(1.3)                              \cr
&\d_x K_1+3A^{-1}\d_x AK_1-A^{-1}\d_x At-\tilde\kappa^{AB}
\tilde\lambda_{AB}=8\pi JA&(1.4)    \cr
&\d_t a=a[-\d_x\beta^1+\f12\alpha(3K_1-t)]&(1.5)          \cr
&\d_t A=-\alpha K_1A+\d_x(\beta^1 A)&(1.6)}               
$$ 
The quantity $K_1$ appearing in these equations is an eigenvalue of the
second fundamental form. Alternatively it may be thought of as an auxiliary
quantity defined in terms of the basic quantities contained in (1.1) by
(1.5). The quantity $\eta_A$ is given by
$$\eta_A=(1/2)\alpha^{-1}Aa^2\tilde g_{AB}\d_x\beta^B\eqno(1.7)$$
while $\tilde\kappa_{AB}$ and $\tilde\lambda_{AB}$ are the tracefree parts of 
the second fundamental forms of the group orbits in spacetime corresponding
to the normal vector to the hypersurface $t$=const. and the normal to the
orbit in the hypersurface $t$=const., respectively. The components $\tilde
g_{AB}$ can be parametrized as follows:
$$\tilde g_{22}=e^W\cosh V,\ \ \ \tilde g_{33}=e^{-W}\cosh V,\ \ \
\tilde g_{23}=\sinh V\eqno(1.8)$$
In terms of $W$ and $V$ the squares of $\tilde\kappa_{AB}$ and 
$\tilde\lambda_{AB}$ have the explicit forms:
$$\eqalignno{
\tilde\lambda_{AB}\tilde\lambda^{AB}&=\f12 A^{-2}(\cosh^2 VW_x^2+V_x^2) 
&(1.9)         \cr
\tilde\kappa_{AB}\tilde\kappa^{AB}&=\f12\alpha^{-2} [\cosh^2 V
(W_t-\beta^1W_x)^2+(V_t-\beta^1 V_x)^2]&(1.10)
}$$
The quantities $\rho$, $J$ and $\tr S$ denote the energy density, the matter
current and three times the mean pressure, respectively.

In the proofs of the existence and non-existence theorems it is important
to have estimates for a solution of these equations on a finite time
interval $(t_1,t_2)$ with $t_2<0$ in terms of the data it induces at some
intermediate time $t_0$. Consider a point where $\alpha$ attains its
maximum on a hypersurface of constant time. The strong energy condition
implies that $\rho+\tr S\ge 0$ and so it follows from equation (1.3) that
$\alpha\le 3/t^2$. In fact this estimate is true, and can be proved in the
same way, without any symmetry assumption on the spacetime. The next step
is to use a generalization of an argument of Malec and \'O Murchadha [3]
to show that if the dominant energy condition holds then
$|K_1|\le 5|t_1|$, $|A^{-2}\d_x A|\le 2|t_1|$. (For details
of this argument see [2].) With these estimates (which depend very much on 
the symmetry assumption) in hand, equations (1.5) and (1.6) can be integrated
in time to give the following bounds for $a$ and $A$.
$$\eqalign{
&a(t_0)\exp (-C|t-t_0|)\le a(t)\le a(t_0)\exp (C|t-t_0|)  \cr
&\sup\{A(t,x),A^{-1}(t,x)\}\le \sup\{\|A(t_0)\|_\infty,
\|A^{-1}(t_0)\|_\infty\}\exp(\exp(C|t-t_0|))}
\eqno(1.11)$$
Along the way it also comes out that $|\d_x\beta^1|$ satisfies the same kind
of upper bound as $a$. For details the reader is once again referred to [2].
These bounds are probably far from optimal but they are sufficient for the
present purposes. The general point is that many aspects of the geometry can
be bounded in terms of $t_1$, $t_2$ and the maximum
and minimum values of $a$ and $A$ on the initial hypersurface $t=t_0$.
Using this information and integrating equation (1.2) in space at each 
fixed time gives a bound for the integral $\int_0^{2\pi}Q(t,x) dx$ where
$$Q=16\pi\rho+2\eta^A\eta_A+\tilde\kappa^{AB}\tilde\kappa_{AB}
+\tilde\lambda^{AB}\tilde\lambda_{AB}   \eqno(1.12)$$
which is uniform in time on the given interval.

\vskip 11 true pt
\centerline{
{\small 2. EXAMPLES OF SINGULARITY FORMATION}}

\vskip 11 true pt

A matter model which is well known for its tendency to form singularities
is dust. In this section a type of example will be presented which shows
that global existence in a CMC time coordinate does not hold for the
Einstein-dust system. The main interest of this is the contrast it provides
with the known positive results for some other matter models which will be 
reviewed
in the next section. To put it another way, this example says something about
the sharpness of the results previously obtained. The intuitive idea behind
the example is very simple. If two dust particles start out close together
and with velocities which are moderate in magnitude and opposite in direction
then they should collide after finite time producing a shell-crossing
singularity. To turn this intuition into a proof we can use the bounds on
various geometric quantities presented in the last section. However these
estimates alone do not suffice. The difficulty is connected with the phrase
\lq in finite time\rq. The problem is that a finite amount of proper time
along the wordline of a particle might correspond to an infinite amount of
coordinate time. This would correspond to the phenomenon known as \lq collapse
of the lapse\rq , where the lapse function $\alpha$ becomes very small. To show
that this does not happen it is necessary to have a positive lower bound on
the lapse function on a finite interval of CMC time. This is the subject of
the following lemma.

\vskip 9 true pt\noindent
LEMMA (supporting the lapse) Consider a solution of the Einstein-matter
equations with $U(1)\times U(1)$ symmetry on a finite interval $(t_1,t_2)$ of 
CMC time with $t_2<0$ and let $t_0$ belong to this time interval. Suppose that
the strong and dominant energy conditions hold. Then there is constant $C$, 
depending only on $t_1$, $t_2$ and the maximum values of $a$, $a^{-1}$, $A$ 
and $A^{-1}$ on the hypersurface $t=t_0$ such that $\alpha^{-1}\le C$.

\vskip 9 true pt

\noindent
Proof. Equation (1.3) can be written in the form
$$\d_x(A\d_x\alpha)=\alpha A^3 P-A^3\eqno(2.1)$$
where $P$ is non-negative. Under the given hypotheses $G=\int_0^{2\pi}
P(x)dx$ is bounded. Consider now a fixed time $t$ and an interval 
$I=[x_1,x_2]$ of values of $x$. Let $L=|x_2-x_1|$ and denote by $F$ and $F'$
the maximum values of $A\alpha$ and $\d_x(A\alpha)$ respectively on the
interval $I$. Now for any $x\in I$
$$(A\alpha)(x)=(A\alpha)(x_1)+\int_{x_1}^x \d_x (A\alpha)(y) dy$$
and the integral can be bounded by $LF'$. Hence:
$$F\le (A\alpha)(x_1)+LF'\eqno(2.2)$$
On the other hand
$$\eqalign{
\d_x(A\alpha)(x)&=(A\d_x\alpha)(x)+(A^{-1}\d_x A (A\alpha))(x)     \cr
&=(A\d_x\alpha)(x_1)+\int_{x_1}^x \d_x(A\d_x\alpha)(y) dy
+(A^{-1}\d_x A (A\alpha))(x)     \cr
&=\d_x(A\alpha)(x_1)-(A^{-1}\d_x A(A\alpha))(x_1)+\int_{x_1}^x 
(\alpha A^3 P-A^3)(y) dy+(A^{-1}\d_x A (A\alpha))(x)}$$
In the last line equation (2.1) has been used. Since $A^{-1}\d_x A$ 
has already been bounded, the second and fourth terms in the final 
expression are bounded by $CF$ for some positive constant $C$. Consider now 
the third term. The second term in the integrand makes a negative 
contribution, while the first can be bounded by $CFP$. Hence:
$$F'\le \d_x(A\alpha)(x_1)+C(1+G)F\eqno(2.3)$$
It follows by substituting (2.3) into (2.2) that
$$F\le (A\alpha)(x_1)+L[(\d_x(A\alpha)(x_1)+C(1+G)F]\eqno(2.4)$$
If $L\le\f12[C(1+G)]^{-1}$ then this implies that
$$F\le 2[(A\alpha)(x_1)+L\d_x (A\alpha)(x_1)]\eqno(2.5)$$
Similarly, under the same restriction on $L$:
$$F'\le 2[\d_x(A\alpha)(x_1)+C(1+G)(A\alpha)(x_1)]\eqno(2.6)$$
Thus we have the inequalities
$$\eqalign{
(A\alpha)(x+L)&\le 2[(A\alpha)(x)+L\d_x (A\alpha)(x)] \cr
\d_x(A\alpha)(x+L)&\le 2[\d_x (A\alpha)(x)+C(1+G)(A\alpha)(x)]}
\eqno(2.7)$$
Now two cases will be considered. The first is where $C(1+G)\le (4\pi)^{-1}$.
Then $L$ can be chosen to be $2\pi$ and it follows that if $A\alpha$ takes its
minimum at a given time at $x_-$ then
$$(A\alpha)(x)\le 2 (A\alpha)(x_-)\eqno(2.8)$$
If $C(1+G)>(4\pi)^{-1}$ then it is possible to divide the circle into a 
number of equal intervals, starting at $x_-$ whose length $L$ satisfies the
desired inequality and whose number does not exceed $C(1+G)$. Let
$z(x)=\max\{(A\alpha)(x),\d_x (A\alpha)(x)\}$. Then
$$z(x+L)\le 2C(1+G)z(x)\eqno(2.9)$$
Applying this on $k$ successive intervals gives
$$z(x+kL)\le 2^k C^k (1+G)^k z(x)\eqno(2.10)$$
On the other hand $A\alpha(x)$ can be bounded by a fixed constant times
$z(x+kL)$ for some $k\le C(1+G)$. There results an estimate of the form
$$(A\alpha)(x)\le [C(1+G)]^{C(1+G)}(A\alpha)(x_-)\eqno(2.11)$$
As discussed in the previous section, the assumptions of the lemma imply
bounds on $A$ and $A^{-1}$. Hence (2.8) and (2.11) imply an estimate of
the form $\|\alpha\|_\infty\le C\alpha_-$, where $\alpha_-$ is the 
minimum value of $\alpha$ at the given time. On the other hand, integrating
equation (2.1) with repect to $x$ shows that:
$$\int_0^{2\pi}A^3(x)dx=\int_0^{2\pi}(\alpha A^3 P)(x) dx
\le C\|\alpha\|_\infty$$
Since $\int_0^{2\pi}A^3(x)dx$ is bounded from below, putting these estimates
together gives the desired lower bound for $\alpha_-$.

\vskip 10pt\noindent
The estimates (2.8) and (2.11) resemble the specialization of Harnack's 
inequality to one space dimension with the difference that the constant
in the inequality only depends on the $L^1$ norm of $P$ rather than its
$L^\infty$ norm.

Next the above lemma will be applied to the Einstein-dust system. The
matter fields are a non-negative function $\mu$, the proper energy
density of the dust, and a unit vector $u^\alpha$, the four-velocity
of the dust particles. The energy-momentum tensor is given by
$T_{\alpha\beta}=\mu u_\alpha u_\beta$ and the equations describing the
dynamics of the matter fields are simply $\nabla_\alpha T^{\alpha\beta}=0$,
which is of course a necessary compatibility condition for the Einstein
equations. The dominant, strong and weak energy conditions are satified by
this matter model. The theorem to be proved is the following:

\vskip 9 true pt\noindent
THEOREM Let $t_0$ be a negative real number and $\epsilon>0$. Then there
exist initial data with constant mean curvature $t_0$ for the Einstein-dust 
system such that in any Cauchy development of these data the CMC foliation
which exists in a neighbourhood of the initial hypersurface cannot be 
extended to values of the mean curvature greater than $t_0+\epsilon$.
Similarly there exist data for which the foliation cannot be extended to
values of the mean curvature less than $t_0-\epsilon$.

\vskip 9 true pt

\noindent
Proof. The initial data to be constructed will have $U(1)\times U(1)$
symmetry and so it is possible to use the form (1.1) of the metric. They will
also be such that the initial velocity of the dust particles is in the 
$x$-direction. Then the velocity can be parametrized by the inner product
$v$ of the four-velocity $u^\alpha$ with the unit vector $A^{-1}\d/\d x$.
The integral curves of $u^\alpha$ are geodesics and hence there are 
conservation laws corresponding to the Killing vectors $\d/\d y^A$. This
means that if the velocity is initially in the $x$-direction, it remains so.
Thus these dust solutions can be described completely by the two functions 
$\mu$ and $v$. 

Initial data for dust spacetimes can be constructed using the conformal 
method. (A general discussion of this method can be found in [4].)
However the situation here is sufficiently simple that we 
will not need all of the general theory. The data constructed will be
such that $\tilde\kappa_{AB}=0$, $\tilde\lambda_{AB}=0$ and $\eta_A=0$. The 
matter quantities occurring in the constraints are related to the matter 
fields $\mu$ and $v$ in the special case of the type of data under 
consideration by:
$$\eqalign{
\rho&=\mu(1+v^2)                  \cr
J&=\mu(1+v^2)^{1/2}v}\eqno(2.12)$$
To construct initial data, choose first the constant value $t=t_0$
for the mean curvature and a non-negative function $\tilde\mu$,
a function  $v$  and a scalar function $\tilde K_1(x)$ on the circle.
The solution of the constraints is sought in the form: 
$$\eqalign{
K_1&=\f13 t+A^{-3}(\tilde K_1-\f13 t)   \cr
\mu&=A^{-4}\tilde\mu}\eqno(2.13)$$
and the Lichnerowicz equation, which is just (1.2) rewritten in terms of 
the rescaled quantities, reads:
$$\d_x^2(A^{1/2})=-\f18A^{5/2}[\f32 A^{-3}(\tilde K_1-\f13t)^2-\f23t^2
+16\pi A^{-4}\tilde\mu (1+v^2)]$$
In terms of the rescaled quantities the equation (1.4) takes the simple
form $\d_x(\tilde K_1-\f13 t)=8\pi\tilde J$, where $\tilde J=\tilde\mu
(1+v^2)^{1/2}v$. This equation can be solved if and only if
$\tilde J$ has integral zero. One way of ensuring that this condition is
satisfied is to impose the symmetry conditions that $\tilde\mu (x)=
\tilde\mu (\pi-x)$ and $v(x)=-v(\pi-x)$. When this equation
is solvable the $L^\infty$
norm of $\tilde K_1-\f13 t$ can be estimated in terms of the $L^1$ norm of
$\tilde J$. The next step is to solve the Lichnerowicz equation and for 
that an idea will be borrowed from the general method, namely that of
using sub- and supersolutions. In order for this to run as smoothly as
possible, assume that $\tilde\mu$ is bounded below by a positive constant $B$.
Then it is possible to find constant sub- and supersolutions, namely
$$\eqalign{
A_-&=t_0^{-1}(24\pi B)^{1/2}                  \cr
A_+&=\max\{t_0^{-1}(48\pi\|\tilde\mu\|_\infty (1+\|v\|_\infty^2))^{1/2},
           t_0^{-2/3}(\f92\|\tilde K-\f13 t\|_\infty^2)^{2/3}\}}\eqno(2.14)$$
These ensure the existence of a solution of the Lichnerowicz equation and give
pointwise estimates for the solution. This allows $A$ and $A^{-1}$ to be 
bounded pointwise in terms of the $L^\infty$ norms of $\tilde K_1-\f13 t$,
$\tilde\mu$ and $v$ and the constants $t_0$ and $B$.

The above provides a possibility of constructing a variety of initial data
in such a way that the quantities entering the hypotheses of the lemma 
above can be easily controlled in terms of the free data. To obtain the
desired example a specific subclass of this data will be considered.
Let $x_1<x_2$ and choose initial data for $v$ such that $v_1=v(x_1)=1$
and $v_2=v(x_2)=-1$. Consider two dust particles which start at time $t_0$
at the points $x_1$ and $x_2$ and with the velocities $v_1$ and $v_2$
respectively. Let the positions and velocities of these particles at
time $t$ be denoted by $x_1(t)$, $x_2(t)$, $v_1(t)$ and $v_2(t)$. These
quantities satisfy the equations:
$$\eqalignno{
dx/dt&=\alpha A^{-1} v/\sqrt{1+v^2}-\beta^1&(2.15)         \cr
dv/dt&=-A^{-1}\alpha'\sqrt{1+v^2}+\alpha Kv&(2.16)}$$
It is elementary to see that on a given finite interval of CMC time bounds
for $v_1$ and $v_2$ can be obtained using Gronwall's inequality. Moreover 
these bounds do not depend on the distance $|x_2(t_0)-x_1(t_0)|$. The idea 
now is to assume that for a family of initial data of this type with 
$|x_2(t_0)-x_1(t_0)|$ tending to zero the corresponding solutions exist 
at least up to and including the time $t_0+\epsilon$ and to show that this 
assumption leads to a contradiction. Note that if $t_0+\epsilon\ge 0$ the 
statement of the theorem is an immediate consequence of the non-existence of 
maximal hypersurfaces. Thus it is assumed in the following that 
$t_0<-\epsilon$.

Let $C$ be a constant greater than 
$5\sup_{t_0<t<t_0+\epsilon}\{(\|A^{-1}\alpha'\|_\infty
+\|\alpha K\|_\infty)\}$. 
By what has been said above we know that the data can be chosen so that a
single constant $C$ works for all data in the family. It will now be shown 
that for the solution evolving from any one of these data
$v_1(t)\ge\f12$ for all times $t\le t_0+\epsilon$ such that 
$t_0\le t\le t_0+C^{-1}$. Let $t_*$ be the largest time in the interval 
$[t_0,t_0+\epsilon]$ such that $v_1(t)\ge\f12$ for all $t$ in the interval 
$[t_0,t_*]$. If $t_*<t_1+C^{-1}$ and $t<t_0+\epsilon$, let $t'$ be the last 
time before $t_*$ that $v_1(t)$ was equal to unity. Now $v(t_*)=\f12$. On the 
other hand it follows by integrating (2.16) from $t'$ to $t_*$ that 
$v(t_*)\ge\f53$. This contradiction shows that in fact either 
$t_*\ge t_0+C^{-1}$ or $t_*=t_0+\epsilon$. This gives the desired conclusion. 
In a similar way it can be shown that $v_2(t)\le -\f12$ for all times 
$t\le t_0+\epsilon$ such that $t\le t_0+C^{-1}$. For convenience of notation, 
let $t_3=\min\{t_0+\epsilon,t_0+C^{-1}\}$. From (2.15):
$$d/dt(x_1-x_2)=\alpha A^{-1}v^1/\sqrt{1+v_1^2}-\alpha A^{-1}v_2/
\sqrt{1+v_2^2}-\beta^1(x_1)+\beta^1(x_2)\eqno(2.17)$$
On the interval $[0,t_3]$ we have a lower bound for $v_1$, an upper bound
for $v_2$ and a crude upper bound for $\sqrt{1+v_1^2}$ and  $\sqrt{1+v_2^2}$.
Moreover, we have a lower bound for $\alpha A^{-1}$. (It is at this point
that the lemma on supporting the lapse is used.) Putting all this together
gives a negative upper bound for the sum of the first and second terms on the
right hand side of (2.17). On the other hand
$$|\beta^1(x_1)-\beta^1(x_2)|\le \|\d_x\beta^1\|_\infty |x_1-x_2|\eqno(2.18)$$
Hence by choosing $|x_1(t_0)-x_2(t_0)|$ small it can be ensured that the
sum of third and fourth terms on the right hand side of (2.17) is negligeable
in comparison with the sum of the first and second terms. Hence it can be
arranged that $x_1(t)-x_2(t)$ tends to zero after a time smaller than $t_3$.
However this contradicts the existence of a regular solution of the 
Einstein-dust equations up to time $t_0+\epsilon$, since in a regular 
solution the world lines of dust particles can never cross. Thus the first
statement of the theorem has been proved. The proof of the second statement
is strictly analogous.

\vskip 11 true pt
\centerline{
{\small 3. EXAMPLES OF GLOBAL REGULARITY}}

\vskip 11 true pt

There are matter models for which the situation is quite different from that
presented in the last section, in that for spacetimes with $U(1)\times U(1)$
symmetry possessing a CMC Cauchy surface global existence in CMC time is 
obtained. In this context \lq global 
existence\rq\ means existence on the longest time interval consistent with
the topological obstruction discussed in Section 1. A global existence theorem
in this sense was proved in [2] for collisionless matter described by the 
Vlasov equation and for the massless scalar field (or more generally wave 
maps). The nature of these two matter models will now be recalled. In the
case of collisionless matter, the matter field is a non-negative real-valued
function $f$ on the space of future-pointing unit timelike vectors in spacetime
(the mass shell) and the equation which describes the dynamics of the matter
is the Vlasov equation. This simply says that the function $f$ is constant
along the curves which are the natural lifts of timelike geodesics to the
mass shell. The energy-momentum tensor at a given spacetime point is obtained 
by integrating the product of $f$ with a suitable weight over the part of the 
mass shell over that point. Details can be found in [5]. In the case of the
massless scalar field, the matter field is a real-valued function $\phi$ on 
spacetime which is supposed to satisfy the wave equation. The energy-momentum
tensor is of the form $T=d\phi\otimes d\phi-\f12 |d\phi |^2 g$. A wave map, 
which is a Lorentzian analogue of the harmonic maps familiar in Riemannian
geometry, is a generalization of this, where the field, instead of taking
values in the real numbers, takes values in an arbitrary complete
Riemannian manifold, known as the target manifold. It is worth to note that 
the vacuum case, i. e. the case $T=0$, is contained in these results as the 
special case $f=0$ or $\phi=0$. The global existence problem is hard even in 
the case of $U(1)\times U(1)$ symmetric vacuum spacetimes. The physics issue 
is whether gravitational waves propagate smoothly for arbitrarily long times
or whether they could develop shocks. It is reasonable to expect that when 
a global existence theorem in CMC time holds, it should not be possible to 
extend the spacetime beyond the region covered by the CMC foliation, while 
maintaining the property that the initial hypersurface be a Cauchy
hypersurface for the extended spacetime. Unfortunately, this statement has
not been proved up to now.

The example of collisionless matter is particularly interesting due to the 
fact that dust solutions can be considered as distributional solutions of the 
Einstein-Vlasov system [6]. They have Dirac $\delta$-function dependence on
the velocity variables. The results already quoted show that approximating
the $\delta$-function by smooth functions in the initial data leads to a
dramatic change in the long-time behaviour.

The basis of the global existence theorem is formed by the estimates mentioned
in Section 1. One then proceeds by bounding higher and higher derivatives of 
the metric and the matter fields on the given finite interval $(t_1,t_2)$. 
When all derivatives have been bounded it follows that the solution can be 
extended to the closed interval. A local existence theorem (which is a 
consequence of standard results on the Cauchy problem and the existence of 
CMC foliations) then allows it to be extended to a longer time interval. 
Finally, consideration of the maximal interval of existence implies the 
desired global theorem. 

The fact that higher derivatives can be bounded is connected to the fact
that the given matter model does not form singularities in a given 
regular spacetime. This condition is violated by dust. The quantity 
$\int_0^{2\pi} \rho(t,x) dx$ is a bounded function of $t$ on the given
time interval, independent of the matter model. On the other hand, it is to 
be expected that $\rho$ can blow up pointwise on a finite time interval in 
the case of dust. For other matter models it could happen that, although the
energy density remains bounded, regularity breaks down at a higher level.
This could result from formation of shocks by the matter. Coming back to
the general case, the PDE problem to be studied is that of a semilinear 
equation of wave map type for $W$ and $V$ coupled to the matter equations. 
The target space of the wave map is the hyperbolic plane. This system is
defined on a curved two-dimensional geometry defined by $\alpha$, $\beta^1$,
$a$ and $A$. What needs to be shown is that the finiteness of a certain
$C^k$ norm of the two-dimensional geometry implies that of a similar norm
of $W$, $V$ and the matter quantities. Conversely, the finiteness of the
latter implies, via equations (1.2)-(1.6), the finiteness of a stronger norm 
of the two-dimensional geometry. This allows higher derivatives of all
quantities of interest to be bounded inductively in favourable cases. The 
details of the argument for bounding low order derivatives depend very much 
on the particular matter model.  

There are similar results for spherically symmetric spacetimes with a
Cauchy hypersurface of topology $S^2\times S^1$. In that case there is
no obstruction to the existence of a maximal hypersurface and all real 
values are attained by the mean curvature [7,8]. Moreover, it can be
shown that the CMC foliation covers the entire spacetime. There are also
some other cases where results are known but they do not involve larger
classes of solutions than those discussed above (for details see [1]).
All the known positive results are for spacetimes with at least two
symmetries, so that the problem effectively reduces to studying a system
in one (or less) space dimension. It seems that even the case with one
symmetry is very difficult, not to mention the general case. However,
investigations into generalizing the results reported here to those
cases are being carried out. It should be noted that symmetries with fixed
points also lead to difficulties so that, for instance, the case of spherical
symmetry on $S^3$ remains open.

Another possible direction for generalizations is to keep the high symmetry
but to relax the assumptions on the matter model. There are two basic types
of matter model to be considered. There are the phenomenological matter
models (e.g. dust, collisionless matter) and the field-theoretic matter
models (e.g. massless scalar field, wave maps). The phenomenological
matter models represent a macroscopic description of matter. They often
have a tendency to form singularities in finite time in a given smooth
spacetime. Dust is a good example. It would probably be possible to prove
an analogue of the theorem of Section 2 for perfect fluids with pressure.
The difference would be that, since the expected singularities are shocks,
the energy density would probably remain finite at the time when the CMC
foliation broke down. To try and get positive results for a fluid, it
would be natural to introduce viscosity. Unfortunately, the concept of
viscous fluids is known to be problematic even in special relativity. A
case where it is difficult to make predictions is that of the Boltzmann
equation. This is because, to the author's knowledge, there is up to now
not even a global existence theorem for classical solutions of the special 
relativistic Boltzmann equation in the case of spatially homogeneous
initial data. 

Field theoretic matter models are supposed to describe matter at a more
fundamental level. They seem less prone to forming singularities in a
given smooth spacetime than phenomenological matter models. They do
have a different kind of problem, which may purely be an incompatibility
with the known techniques, rather than an essential difficulty. The 
problem is that in certain steps of the proofs one would like to have 
the non-negative pressures condition, i.e. the condition that $T(X,X)\ge 0$
for all spacelike vectors $X$. This condition is almost never satisfied 
by field-theoretic matter models. It is not even satisfied by the massless
scalar field. However in the case of $U(1)\times U(1)$ symmetry (or
spherical symmetry) it suffices that $T(X,X)\ge 0$ for spacelike vectors $X$ 
orthogonal to the orbits of the symmetry group. This latter condition is 
satisfied by the massless scalar field and, more generally, by wave maps.
It is not satisfied by an electromagnetic field, a Yang-Mills field or a
massive scalar field. This also has the inconvenience that the positive 
results on existence of CMC hypersurfaces in the case of collisionless
matter do not obviously extend to the case of charged collisionless matter,
coupled to an electromagnetic field. An argument used in [2], and which
does not require non-negative pressures might allow one to circumvent that
difficulty in the $U(1)\times U(1)$ symmetric case. However, it does not,
as it stands, apply to the spherically symmetric case. The massive scalar
field is even worse, since it does not satisfy the strong energy condition,
which is the standard condition used to ensure uniqueness of CMC 
hypersurfaces. Perhaps some use can be made of the fact that the condition
is satisfied in an average sense ([9], p. 95). It would be desirable to have 
more flexible techniques and clearly a lot remains to be learned in this area. 

To conclude, it is in order to make some remarks which put the question of 
the existence of global foliations by CMC hypersurfaces into a wider context. 
One of the most important mathematical open questions in general relativity 
is the cosmic censorship hypothesis of Penrose. It is closely related to the 
issue of the global behaviour of solutions of the Einstein equations
corresponding to general initial data. In [10] Eardley and Moncrief suggested 
that CMC hypersurfaces could be useful in trying to confirm this hypothesis. 
More generally, results of the kind discussed in this paper represent 
knowledge about long-time existence of solutions of the Einstein equations. 
This is not by itself enough to say something about cosmic censorship. For 
that one would not only need to know long time existence but also need 
detailed information on the asymptotic behaviour of the solutions.

\vskip 11 true pt
\centerline{
{\small REFERENCES}}

\vskip 11 true pt

{\small
\noindent
1. RENDALL A. D., Constant mean curvature foliations in cosmological 
spacetimes, {\smallit Helv. Phys. Acta} (to appear).
\next
2. RENDALL A. D., Existence of constant mean curvature foliations in
spacetimes with two-dimensional local symmetry, Preprint AEI 009, 
gr-qc/9605022.
\next
3. MALEC E. \& \'O MURCHADHA N., Optical scalars and singularity
avoidance in spherical spacetimes, {\smallit Phys. Rev. D} {\smallbf 50}, 
6033-6036 (1994).
\next
4. CHOQUET-BRUHAT Y. \& YORK J. W., The Cauchy problem, in {\smallit General
Relativity and Gravitation, Vol. 1} (Edited by A. HELD), pp. 99-172.
Plenum, New York (1980).
\next
5. RENDALL A. D., An introduction to the Einstein-Vlasov system, Preprint 
AEI 005, gr-qc/9604001.
\next
6. RENDALL A. D., Cosmic censorship and the Vlasov equation, {\smallit Class. 
Quantum Grav.} {\smallbf 9}, L99-L104 (1992).
\next
7. RENDALL A. D., Crushing singularities in spacetimes with spherical, plane 
and hyperbolic symmetry. {\smallit Class. Quantum Grav.} {\smallbf 12}, 
1517-1533 (1995).
\next
8. BURNETT G. A. \& RENDALL A. D., Existence of maximal hypersurfaces in some 
spherically symmetric spacetimes. {\smallit Class. Quantum Grav.} 
{\smallbf 13}, 111-123 (1996).  
\next
9. HAWKING S. W. \& ELLIS, G. F. R., The large scale structure of space-time.
Cambridge University Press, Cambridge (1973).
\next
10. EARDLEY D. \& MONCRIEF V., The global existence problem and cosmic
censorship in general relativity. {\smallit Gen. Rel. Grav.} {\smallbf 13}, 
887-892 (1981).}

\end